\documentclass{PoS}

\title{
{\small{DESY 13--117, DO-TH 13/19, SFB/CPP-13-48, LPN 13-050}}
\\
Heavy-quark production in deep-inelastic scattering}

\ShortTitle{Heavy-quarks in DIS}

\author{\speaker{Sergey Alekhin}\thanks
{This work has been supported in part 
by Helmholtz Gemeinschaft under contract VH-HA-101 ({\it Alliance Physics at the Terascale}), 
DFG Sonderforschungsbereich/Transregio~9 and by the European Commission through contract      
PITN-GA-2010-264564 ({\it LHCPhenoNet})}\\
        DESY, Platanenallee 6, D--15738 Zeuthen, Germany\\
Institute for High Energy Physics,142281 Protvino, Moscow region, Russia\\
        E-mail: \email{sergey.alekhin@desy.de}}

\author{Johannes Bl\"umlein\\
        DESY, Platanenallee 6, D--15738 Zeuthen, Germany\\
        E-mail: \email{johannes.bluemlein@desy.de}}

\author{Sven-Olaf Moch\\
        DESY, Platanenallee 6, D--15738 Zeuthen, Germany\\
II. Institut f\"ur Theoretische Physik, Universit\"at Hamburg
    Luruper Chaussee 149, D-22761 Hamburg, Germany\\
        E-mail: \email{sven-olaf.moch@desy.de}}

\abstract{We report recent experimental and theoretical progress 
          concerning the heavy-quark electro-production in the context 
          of the ABM11 parton distribution function (PDF) fit. 
          In the updated ABM11 analysis, 
          including the recent combined HERA charm data, 
          the ${\overline{\rm MS}}$-values of the $c$-quark mass 
          $m_c(m_c)=1.24 \pm 0.03
          (\rm{exp})\,^{+0.03}_{-0.02}(\rm{scale})\,^{+0.00}_{-0.07}
          (\rm{th})$ and $m_c(m_c)=1.15\, \pm 0.04
          (\rm{exp})\,^{+0.04}_{-0.00} (\rm{scale})$ are determined at NNLO 
          and NLO, respectively. The values of $m_c$ obtained are 
          compared to other determinations including the ones based on the
          various variable-flavor-number (VFN) scheme prescriptions.
          The VFN scheme uncertainties related to the matching of 
          the 4(5)-flavor PDFs with the 3(4)-flavor ones are discussed.}

\FullConference{XXI International Workshop on Deep-Inelastic Scattering and Related Subject -DIS2013,\\
		22-26 April 2013\\
		Marseilles,France}

\begin{document}

The $c$- and $b$-quarks provide an important experimental and 
phenomenological tool to study the nucleon structure. 
Experimental separation of the heavy quarks in the final state is facilitated
due to their relatively large masses. On the other hand, 
since the masses $m_{c,b}\gg \Lambda_{QCD}$, with $\Lambda_{QCD}$ stands for the
QCD  scale, the Wilson coefficients for heavy-quark production can be calculated 
within perturbative QCD. The study of heavy-quark production in the deep-inelastic scattering 
(DIS) 
process has been started in the fixed-target experiments. However, only 
at the energies available at HERA  it gives a substantial contribution to 
the inclusive structure functions (SFs). Through the photon-gluon 
fusion mechanism the semi-inclusive SFs of the $c$- and $b$-quark DIS production are 
directly connected to the gluon distribution. Therefore they  
are customary employed in the parton-distribution function (PDF) analyses 
as an additional constraint on the small-$x$ behavior of the gluon distribution. The main 
theoretical 
difficulty arising in this context is related to the emergence of  
two hard scales, given by the quark mass and the DIS momentum transfer $Q^2$. 
At $Q^2 \gg m_{c,b}^2$ power corrections of $O(m_Q^2/Q^2)$ may be neglected 
and the 
massive Wilson coefficients can be presented as a convolution of the 
massless coefficients with the massive operator matrix 
elements (OMEs)~\cite{Buza:1996wv,Aivazis:1993pi}.
This approach serves a basis of the variable-flavor-number (VFN) 
scheme trying to overcome the difficulties of the full massive calculations. 
However, the asymptotic regime poorly overlaps with the kinematics of 
the present data at HERA, which abundantly populate the low-$Q^2$ region. 
In contrast, the fixed-flavor-number (FFN) scheme provides an accurate 
treatment of the mass effects at threshold. 
Moreover, this scheme has demonstrated very good agreement with the existing DIS 
data up to the largest values of $Q^2$~\cite{Alekhin:2012ig}. 
In the following we describe 
the impact of the new charm-production data on the ABM PDF fit~\cite{Alekhin:2012ig}
related to the recent theoretical progress in the FFN scheme
calculations. We report the value of $m_c$ extracted from the DIS data
alongside with the analysis of its uncertainty 
and discuss additional uncertainties on $m_c(m_c)$ and  strong coupling 
constant $\alpha_s$ emerging in the VFN scheme. 

%---------------------------------------------------------------------------------------
\begin{figure}[ht]
  \centering
  \includegraphics[width=0.8\textwidth,height=3.3in]{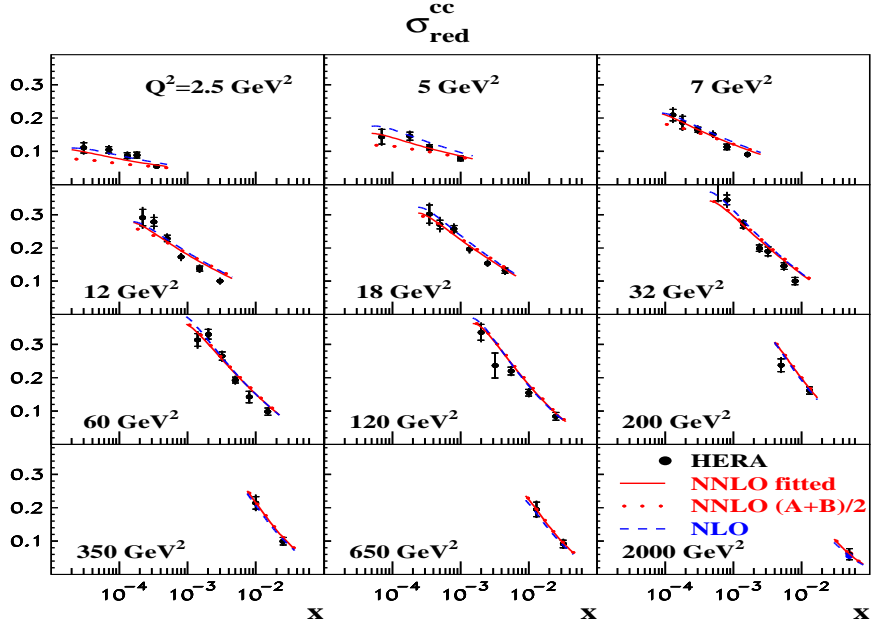}
\caption{\protect\label{fig:herac}                                              
        \small The combined HERA data on the open 
  charm production~\protect\cite{Abramowicz:1900rp} versus $x$ at different
  values of $Q^2$ in comparison with the analysis of~\protect\cite{Alekhin:2012vu} at   
  NLO (dashed line) and NNLO (solid line) together with 
  a fit variant based on the option (A+B)/2 of the NNLO Wilson
  coefficients of Ref.~\protect\cite{Kawamura:2012cr} (dotted line);
  from Ref.~\protect\cite{Alekhin:2012vu}.
} 
\end{figure}
%---------------------------------------------------------------------------------------

The recent version of the ABM PDF fit~\cite{Alekhin:2012ig} is based on the running-mass 
definition of the 
massive Wilson coefficients~\cite{Alekhin:2010sv} with  
the values of $m_{c,b}$ fixed at the PDG 
values~\cite{Beringer:1900zz}. However, $m_{c}$ can be 
also determined from the 
H1 data on charm production~\cite{Alekhin:2012un} and the constraint 
on $m_{c}$ coming from the combined HERA charm data~\cite{Abramowicz:1900rp} 
turns out to be even more substantial. Using advantages of that experimental input
we perform a variant of the ABM PDF fit with the 
combined HERA data added and the value of $m_{c}$ fitted simultaneously 
with the value of $\alpha_s$ and the PDF 
parameters~\cite{Alekhin:2012vu}.
A model of main massive Wilson coefficients employed in this fit 
has been derived in Ref.~\cite{Kawamura:2012cr}
as a combination of the threshold resummation 
results~\cite{LoPresti:2011zz} with the high-energy asymptotics of the 
DIS structure functions~\cite{Catani:1990eg}.
These two regimes are matched
using the available Mellin moments of the 
NNLO massive OMEs and functions~\cite{
Bierenbaum:2008yu,Bierenbaum:2009mv}. 
Furthermore, the calculations are performed within the  
running-mass definition providing improved perturbative 
convergence of the result~\cite{Alekhin:2010sv}. 
To quantify the uncertainty in the 
approximate NNLO coefficients obtained in this way 
two options of these coefficients, A and B, are provided 
in Ref.~\cite{Kawamura:2012cr}. In our analysis we employ a linear 
combination of these options with an interpolation parameter $d_N$
fitted simultaneously with the other fit parameters. The value of $d_N=-0.1$ found 
corresponds to the coefficient shape close to option A. 
The option B is disfavored by the HERA charm data~\cite{Abramowicz:1900rp},
cf. Fig.~\ref{fig:herac}, 
with $\chi^2/NDP=115/52$ obtained in the variant of the fit with this shape, 
where $NDP$ stands for the number of data points. Therefore we quantify 
uncertainties due to the massive NNLO coefficients by the  difference between 
the results obtained with the value of $d_N=-0.1$ preferred by the data
and $d_N=0.5$, corresponding to the average of the options A and B.

\begin{figure}[ht]
\center
  \includegraphics[width=0.8\textwidth,height=2.2in]{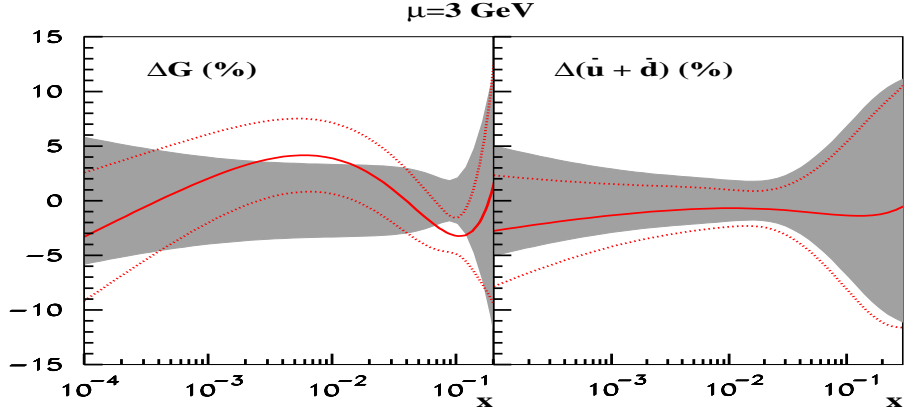}
\caption{\label{fig:pdfs}
  The relative change in the NNLO gluon (left) and non-strange sea
  (right) distributions obtained in the present analysis
  with respect to the ABM11 PDFs (solid lines). The relative uncertainties
  in the PDFs are displayed for comparison (shaded area: ABM11, dotted lines:
  present analysis).
}
\end{figure}

The PDFs obtained in this version of the ABM fit including the 
HERA charm data are compared with those of ABM11 in Fig.~\ref{fig:pdfs}.
The change in the sea quark distribution is moderate and the change 
in the valence region is even smaller. At the same time  
the gluon distribution changes by $1\sigma$ in places both due to 
impact of the new experimental and the theoretical improvements in the 
heavy-quark treatment. 
The $\overline{\rm MS}$-values of the $c$-quark mass obtained in our analysis are 
\begin{eqnarray}
  \label{eq:mcres-nlo}
  m_c(m_c) \,\,=&
  1.15\, \pm 0.04 (\rm{exp})\,^{+0.04}_{-0.00} (\rm{scale})
  \hspace*{30mm}
  &{\rm NLO} 
  \, ,
  \\
  \label{eq:mcres-nnlo}
  m_c(m_c) \,\,=&
  1.24\, \pm 0.03 (\rm{exp})\,^{+0.03}_{-0.02} (\rm{scale})\,^{+0.00}_{-0.07} (\rm{th}),
  \hspace*{14mm}
  &{\rm NNLO_{\rm{approx}}}
  \, ,
\end{eqnarray}
at NLO and NNLO, respectively. The NLO value of 
$m_c(m_c)=1.26\pm0.05~(\rm{exp})~{\rm GeV}$ extracted form the HERA 
data only~\cite{Abramowicz:1900rp} is somewhat bigger than 
ours in Eq.~(\ref{eq:mcres-nlo}). The difference between these two 
determinations was found to appear mainly due to the selection of the data 
employed in the analysis, cf.~\cite{Alekhin:2012vu} for details. 
%---------------------------------------------------------------------------------------
%---------------------------------------------------------------------------------------
The theoretical errors, 
Eq.~(\ref{eq:mcres-nlo},\ref{eq:mcres-nnlo}), emerge due to the 
factorization scale variation by a factor of $1/2$ and $2$ around 
the nominal value of $\sqrt{Q^2 + \kappa m_c^2}$ and due to the NNLO 
coefficient shape uncertainty~\footnote{The factor of $\kappa$ is  
4 and 1 for the neutral- and charged-current cases, respectively.}. 
The NNLO central value is comparable with the one obtained 
from the $e^+e^-$ data and the total error is competitive with the 
world average~\cite{Beringer:1900zz}.

In comparison to the FFN scheme the VFN scheme brings in two additional 
uncertainty sources. The first is related to modeling of 
the low-$Q^2$ region, which is necessary to provide a reasonable 
behavior of the VFN scheme in the kinematic region of the present DIS data. 
This uncertainty was in particular quantified by the extraction of $m_c$ within 
various prescriptions of the VFN scheme, including ACOT-full, S-ACOT-$\chi$,
RT-standard, and RT-optimized prescriptions. While the quality of the 
data obtained with different 
prescriptions is similar, the value of $m_c$ preferred by the data differs by 
$\pm200~{\rm MeV}$~\cite{Abramowicz:1900rp}. 
This estimate is comparable with the uncertainty in $m_c$ due to 
variation of the S-ACOT-$\chi$ prescription parameters~\cite{Gao:2013wwa}. 
The second source of uncertainty is related to the generation of 
the 4(5)-flavor PDFs. They are commonly 
matched with the 3(4)-flavor ones at the scale of 
$\mu_0=m_c(m_b)$~\footnote{Note, at the scale 
of $m_b$ the charm mass effects cannot be fully neglected.}.
%----------------------------------------------------------------------------------------
\begin{figure}[ht]
  \centering
  \includegraphics[width=0.8\textwidth,height=2.0in]{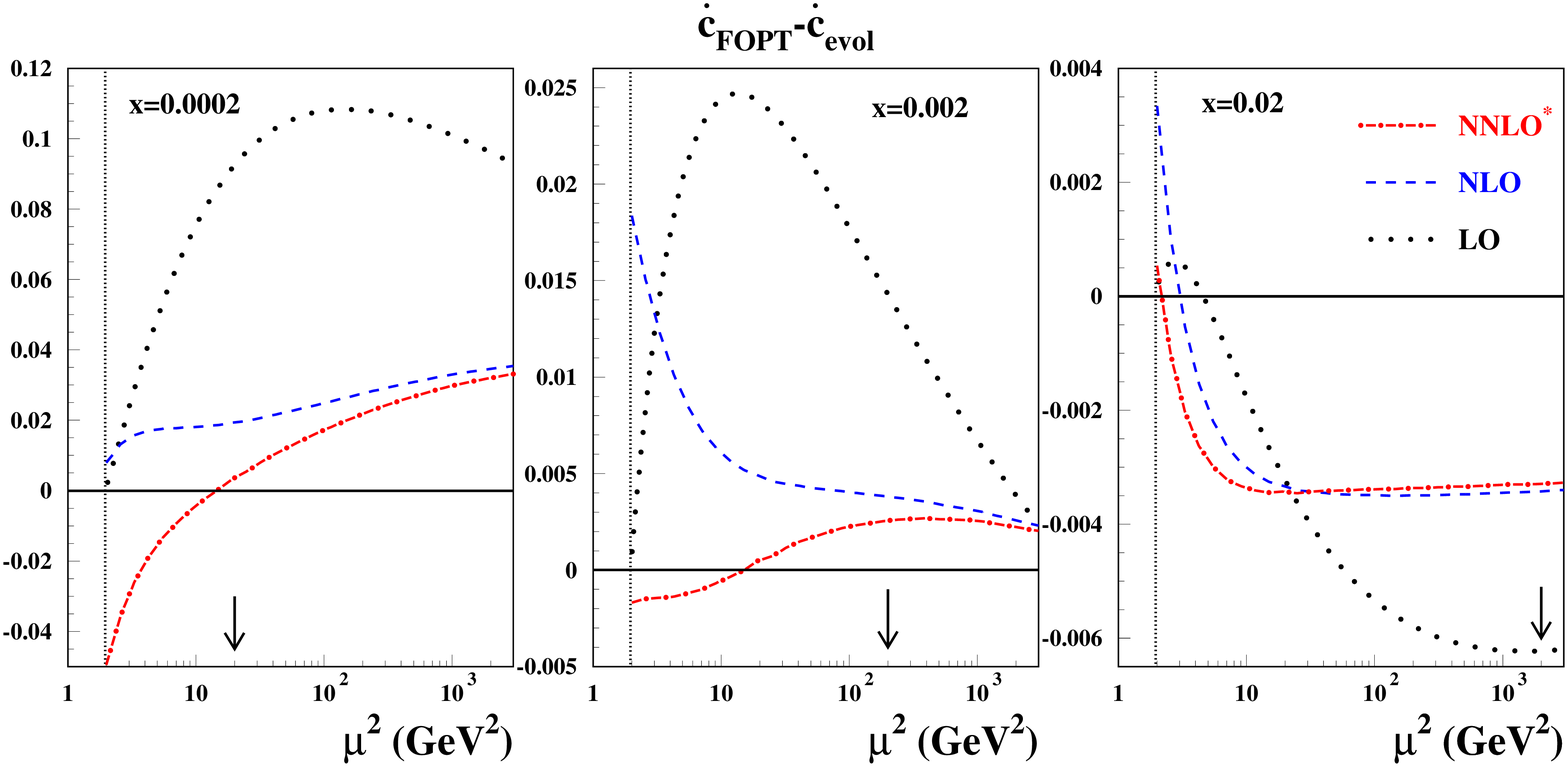}
\caption{\label{fig:pdfder} \small The difference between the 
$c$-quark PDFs derivatives 
$\dot{c}(x,\mu^2)\equiv \frac{dc(x,\mu^2)}{d\ln\mu^2}$ calculated with 
the FOPT matching condition and with the massless 4-flavor evolution 
starting at the matching point $\mu_0=m_c=1.4~{\rm GeV}$ versus the
factorization scale $\mu^2$ at different values of $x$ in the LO, 
NLO, and NNLO* approximations. The arrows display the upper margin of the 
HERA collider kinematics with the collision 
c.m.s. energy squared $s=10^5~{\rm GeV}^2$ and the vertical lines correspond to
the matching point position $\mu_0$.}
\end{figure}
%----------------------------------------------------------------------------------------
This 
is an arbitrary choice of course and the variation of the matching point $\mu_0$ in a wide 
range
is allowed in principle. Further, the 4(5)-flavor PDF obtained in this 
way are evolved starting from the scale $\mu_0$ using massless
evolution kernels. In the NNLO case this cannot be performed 
consistently since the NNLO OMEs are not yet fully 
known~\footnote{For progress in this field, cf.~\cite{Ablinger:2012ej}.}.
In practice, the the NNLO evolution is commonly combined with the 
NLO matching at $\mu_0$  arriving at an approximation 
called NNLO$^*$ in the following. 
The theoretical uncertainties in the latter are illustrated by
comparison of the $c$-quark distributions $c(x, \mu^2)$ generated at NNLO$^*$ to
the NLO ones, which are  generated using the NLO matching in 
combination with the NLO evolution.
We consider the derivative of $c(x,\mu^2)$ w.r.t. the factorization scale $\mu^2$
and take the difference of this derivative with the one calculated in 
fixed-order-perturbative theory (FOPT) employing the massive OMEs
to produce $c(x,\mu^2)$ at all values of $\mu^2$. This representation
allows to check the impact of the $\ln\mu^2$-resummation  
manifesting in the PDF evolution at large $\mu^2$. This resummation 
reproduces 
the higher-order correction effects in part. Therefore the difference 
between the FOPT and evolved PDFs vanishes with perturbative order. 
At NLO and NNLO$^*$ the resummation effects are numerically significant 
at $x\lesssim 0.0001$ and at $\mu^2$ outside of the HERA kinematics only, 
cf. Fig.~\ref{fig:pdfder}. In particular this signals that the FFN scheme can 
be reliably used in the NNLO analysis of the HERA data. 
At the same time the uncertainty in the NNLO$^*$ approximation of the 
VFN scheme is localized at small $\mu^2$ well covered by the HERA data.
The impact of this uncertainty combined with the variation of the matching 
point $\mu_0^2$ within the range of $1.2\div 1.5~{\rm GeV}$ on 
$\alpha_s(M_Z)$ is estimated as $\pm 0.001$ for the VFN variant of the ABM11 
fit. In combination with the uncertainty due to the low-$Q^2$ modeling
this makes the VFN scheme uncompetitive with the FFN one in the 
precision determination of $\alpha_s$.

\end{document}